\documentclass[twocolumn,showpacs,prl]{revtex4}
\usepackage{amsmath,amssymb,graphicx}
\usepackage{color,ulem}

\begin{document}

\title{Dynamical affinity in opinion dynamics modelling}

\author{Franco Bagnoli$^{1}$\thanks{franco.bagnoli@unifi.it},
Timoteo Carletti$^{2}$\thanks{timoteo.carletti@fundp.ac.be},
Duccio Fanelli$^{3}$\thanks{duccio.fanelli@manchester.ac.uk},
Alessio Guarino$^{4}$\thanks{guarino@upf.pf}, 
Andrea Guazzini$^{1}$\thanks{andrea.guazzini@unifi.it}}

\affiliation{ 1. Dipartimento di Energetica and CSDC, Universit\`a di Firenze, and INFN, via S. Marta, 3, 50139 Firenze, Italy\\
  2. D\'epartement de Math\'ematique, Universit\'e Notre Dame de la Paix, 8 rempart de la vierge B5000 Namur, Belgium\\ 
  3.Theoretical Physics, School of Physics and Astronomy, University of Manchester, M13 9PL, Manchester, United Kingdom\\ 
  4. Universit\'e de la Polyn\'esie Francaise, BP 6570 Faa'a, 98702, French Polynesia  
  } \date{\today}

\begin{abstract}
We here propose a  model to simulate the process of opinion
formation, which accounts for the mutual affinity 
between interacting agents. Opinion and affinity evolve self-consistently,
manifesting a highly non trivial interplay. 
A continuous transition is found between single and multiple opinion states. Fractal dimension and signature of critical  
behaviour are also reported. A rich phenomenology is presented and discussed
with reference to corresponding  
psychological implications.
\end{abstract}


\maketitle

The paradigms of complex systems are nowadays being applied to an ample
spectrum of interdisciplinary problems,  
ranging from molecular biology to social sciences. The challenge is to model the dynamical evolution of 
an ensemble made of interacting, microscopic, constituents and infer the emergence of collective, macroscopic, behaviours that are 
then eventually accessible for direct experimental inspection. Statistical mechanics and non--linear physics provide quantitative 
tools to elucidate the key mechanisms underlying the phenomena under scrutiny, often resulting in novel interpretative frameworks.
Agent-based computational models have been widely employed for simulating complex adaptive systems, in particular with 
reference to sociophysics applications. Within this context, opinion dynamics has recently attracted a growing interests
clearly testified by the vast production of specialised contributions \cite{opinion}. Peculiar aspects of its intrinsic dynamics make 
opinion formation a rich field of  analysis where self-organisation, clustering and polarisation occur.  

Opinion dynamics models can be ideally grouped into two large classes. The first deals with binary opinions: 
agents behave similarly to magnetic spins and just two states are allowed (up or down) \cite{Sznajd}. Here social actors update their 
opinions driven by a social influence pressure, which often translates into a majority rule. Alternatively, opinions can be 
schematised with continuous variables, the latter being dynamically evolved as a result of subsequent interactions among 
individuals. In the celebrated Deffuant et al. \cite{Deffuant} model agents adjust their opinion as a results of random binary encounters 
whenever their difference in opinion is below a given threshold. The rationale behind the threshold ansatz reflects
humans' natural tendency to avoid conflicting interests and consequently
ignore the perception of incompatibility between two 
distant cognitions. In this respect, the threshold value measures the average
{\it openness of mind} of the community.  

In real life, the difference in opinion on a debated issue is indeed playing a
crucial role. However, the actual outcome of  
an hypothetic binary interactions also relies on a number of other factors,
which supposedly relate to the quality of the 
inter-personal relationships. Mutual {\it affinity} condensates in fact past
interactions' history and contributes to select preferential  
interlocutors for future discussions. Previous attempts aimed at incorporating
this effect resulted in static  
descriptions, which deliberately disregarded affinity's self-consistent
evolution \cite{Nowak}. In this Letter we take one step forward by 
proposing a novel formulation where the affinity is dynamically
coupled to the opinion, and consequently updated in
time. Moreover, affinity translates in a social 
distance, a concept that is here introduced to drive preferential interactions 
between affine individuals. Macroscopically, the system  is shown to asymptotically organise in
clusters of agents sharing a common opinion, whose number depends on the choice of the parameters involved. Interestingly, 
a continuous transition is identified that separates the mono-clustered from the fragmented phase. Scaling laws are also found and 
their implications discussed. Most importantly, our proposed theoretical scenario captures the so-called {\it cognitive dissonance} phenomenon, 
a qualitatively well documented theory in psychology pioneered by Leon Festinger in 1956 \cite{festinger}. 

Consider a population of $N$ agents, each bearing
at time $t$ a scalar opinion $O_i^{t} \in
  [0,1]$. Moreover, let us introduce the $N \times N$
time dependent matrix 
${\bf \alpha}^{t}$, whose elements $\alpha_{ij}^{t}$ are bound to the
    interval $[0,1]$. Such elements specify the affinity of individual
    $i$ vs. $j$, larger  
numbers being associated to more trustable relationships. Both the
opinions vector and the affinity matrix are randomly initialized at
time $t=0$.  
At each time step $t$, two agents, say $i$ and $j$, are selected
according to a strategy that we shall elucidate in the forthcoming
discussion.  
They interact and update their characteristics according to the
following recipe \footnote{The evolution of the quantities $O_j(t)$
  and $\alpha_{ij}(t)$ 
is straightforwardly obtained by switching the labels $i$ and $j$ in
the equations.}:  

\begin{eqnarray}
O_i^{t+1} &=& O_i^{t}- \mu \Delta O_{ij}^{t} \Gamma_1\left(\alpha^t_{ij}\right) \label{opinion} \\
\alpha_{ij}^{t+1} &=& \alpha_{ij}^{t} + \alpha_{ij}^{t}
      [1-\alpha_{ij}^{t}] \Gamma_2 \left(\Delta O_{ij}\right) 
\label{alpha}
\end{eqnarray}

where the functions $\Gamma_1$ and $\Gamma_2$ respectively read:
\begin{eqnarray}
\Gamma_1 \left(\alpha^t_{ij}\right)&=& \frac{1}{2}\left[ \tanh (\beta_1
  (\alpha_{ij}^{t}-\alpha_c)) + 1 \right] \\ 
\Gamma_2 \left(\Delta O_{ij}\right)&=& -\tanh(\beta_2 (|\Delta O_{ij}^{t}| - \Delta O_c))
\end{eqnarray}

Here, $\Delta O_{ij}^{t} = O_{i}^{t} - O_{j}^{t}$, while $\alpha_c$, $\Delta O_c$ are constant
  parameters. For the sake of simplicity we shall  
consider the limit $\beta_{1,2} \rightarrow \infty$, which practically
  amounts to replace the hyperbolic tangent, with a simpler  
step function profile. 
Within this working
  assumption, the function $\Gamma_{1}$ is 0 or 1, while $\Gamma_{2}$ ranges from -1 to
  1, depending on the value of the arguments. $\Gamma_{1}$  and $\Gamma_{2}$    
act therefore as effective switchers. Notice that, for $\alpha_c
  \rightarrow 0$, equation (\ref{opinion}) reduces to Deffuant et
  al. scheme  
\cite{Deffuant}. To clarify the ideas inspiring our proposed formulation, we shall focus on specific examples. First, suppose two subjects meet and 
imagine they confront their opinions, assumed to be divergent ($|\Delta O_{ij}| \simeq 1$). According to Deffuant's model, when the disagreement exceeds a
fixed threshold, the agents simply stick to their positions. Conversely, in the present case, the interaction can still result in a modification of 
each other beliefs, provided the mutual affinity $\alpha_{ij}^{t}$ is larger than the reference value $\alpha_c$. In other words, individual 
exposed to conflicting thoughts, have to
  resolve such dissonance in opinion by taking one of two opposite
  actions: If $\alpha_{ij}^{t}<\alpha_c$, the agent ignores the  
contradictory information, which is therefore not assimilated; when instead the opinion is coming from a trustable source ($\alpha_{ij}^{t}>\alpha_c$),
the agent is naturally inclined to seek consistence among the cognitions, and consequently adjust its belief. The mechanism here outlined is  
part of Festinger's cognitive dissonance theory \cite{festinger}: contradicting cognitions drive the mind to modify existing beliefs to reduce the
amount of dissonance (conflict) between cognitions, thus  removing the feeling of uncomfortable tension. The scalar $\alpha_{ij}$   
schematically accounts for a larger number of hidden variables (personality, attitudes, behaviours,..), which are non trivially integrated in an
abstract affinity concept. Notice that the matrix ${\bf \alpha}^{t}$ is non symmetric: hence, following a random encounter between two dissonant 
agents, one could eventually update his opinion, the other still keeping his own view. A dual mechanism governs the self-consistent 
evolution for the affinity elements, see equation (\ref{alpha}). If two people gather together and discover to share common interests 
($|\Delta O_{ij}^{t}| < \Delta O_c$) they will increase their mutual affinity ($\alpha_{ij}^{t} \rightarrow 1$). On the contrary, the fact of occasionally facing 
different viewpoints ($|\Delta O_{ij}^{t}| > \Delta O_c$), translates
  in a reduction of the affinity indicator ($\alpha_{ij}^{t}
  \rightarrow 0$). The logistic contribution 
in equation (\ref{alpha}) confines $\alpha_{ij}^{t}$ in the interval $[0,1]$. Moreover, it maximises the change in affinity for pairs
with $\alpha_{ij}^{t} \simeq 0.5$, corresponding to agents which have not come often in contact. Couples with 
$\alpha_{ij}^{t} \simeq 1$ (resp. $0$) have already formed their mind and, as expected, behave more conservatively.

Before turning to illustrate the result of our investigations, we shall discuss the selection rule here implemented. First the agent $i$ is randomly extracted, with 
uniform probability. Then we introduce a new quantity $d_{ij}$,
hereafter termed {\it social distance}, defined as  
\footnote{The affinity can mitigate the difference in opinion, thus
  determining the degree of social similarity of two individuals.  
This observation translates into the analytical form here postulated
  for  $d_{ij}^t$.}   
\begin{eqnarray}
d_{ij}^t &=& \Delta O_{ij}^{t} (1-\alpha_{ij}^{t}) \qquad j=1,...,N \qquad j \ne i.
\label{social_distance}
\end{eqnarray}

The smaller the value of $d_{ij}^t$ the closer the agent $j$ to $i$,
both in term of affinity and opinion. A random, normally distributed,
vector $\eta_j(0,\sigma)$ of size $N-1$ is subsequently generated, with mean
zero and variance $\sigma$. The social distance is then modified  
into the new social metric 
$D_{ij}^\eta = d_{ij}^t + \eta_j(0,\sigma)$.  
Finally, the agent $j$
which is closer to $i$ with respect to the measure 
$D_{ij}^\eta$  is selected  
for interaction. The additive random perturbation $\eta$ is hence acting on a fictictious 1D manifold, 
which is introduced to define the pseudo-particle (agent) interaction on the basis of a nearest neighbors 
selection mechanism. $\eta$ is thus formally equivalent to 
a {\it thermal noise} \cite{guarino}. Based on this analogy, $\sigma$ is here baptized {\it social
  temperature} and set the level of mixing in the community.  
Notably, for any value of $\sigma$, it is indeed possible that agents
initially distant in the unperturbed social space $d_{ij}^t$ 
 mutually interact: their chances to meet increase for larger values of the social temperature.  

\begin{figure}[htbp]
\centering
\includegraphics[width=8cm]{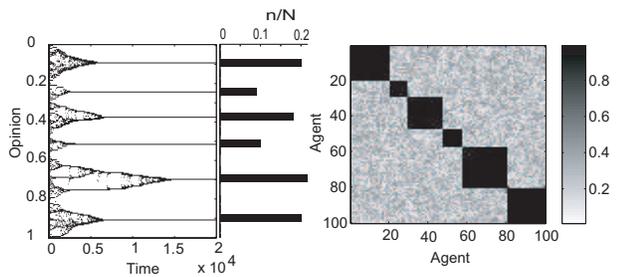}
\caption{Left: Typical evolution of the opinion versus time, i.e. number of
  iterations.  
Right plot: Final affinity matrix. Here $\sigma=0.02$, $\Delta O_c =0.5$, $\alpha_c=0.5$. 
Initial opinion are (random) uniformly distributed within the interval
$[0,1]$.   
$\alpha_{ij}^0$ is initialised with uniform (random) values between $0$ an
$0.5$. Here, $\beta_1=\beta_2=1000$.} 
\label{fig1}
\end{figure}

Numerical simulations are performed and the dynamical evolution of the system monitored. Qualitatively, asymptotic clusters of opinion are formed, 
whose number depends on the parameters involved. The individuals that reach a consensus on the question under debate are also characterised by  
large values of their reciprocal affinity, as clearly displayed in Figure \ref{fig1}. The final scenario results from a non trivial dynamical interplay between 
opinion and affinity: the various agglomerations are hence different in size and, centred around distinct opinion values, which cannot be predicted a priori. 
The dynamics is therefore significantly more rich, and far more realistic, than that arising within the framework of the original Deffuant et al. scheme 
\cite{Deffuant}, where cluster number and average opinions are simply related to the threshold amount. Notice that, in our model, the affinity enters both the 
selection rule and the actual dynamics, these ingrendients being crucial to
reproduce the observed self-organization. 

\begin{figure}[htbp]
\centering
\includegraphics[width=8cm]{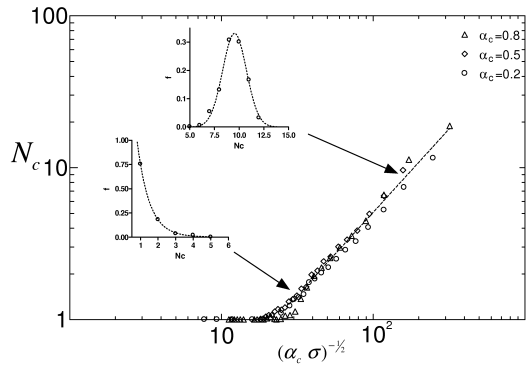}
\caption{Average number of clusters as function of the rescaled quantity $(\sigma \alpha_c)^{-1/2}$. A phase transition is found at 
$(\sigma \alpha_c)^{-1/2} \simeq 20$. Above the transition, histograms of the number of clusters are computed and enclosed as insets 
in the main frame: symbols refer to the numerics, solid lines are fitted interpolation. Here, $\Delta O_c = 0.5$. The variables $O_i^0$ and $\alpha_{ij}^0$ are 
initialised as described in the caption of Figure \ref{fig1}.}
\label{fig2}
\end{figure}

To gain quantitative insight into the process of opinion formation, we run several simulations relative to different initial realizations and 
recorded the final (averaged) number of clusters, $N_c$, as function of the social temperature $\sigma$, for different values of the critical parameter 
$\alpha_c$. Results of the numerics are reported in Figure \ref{fig2}. All the curves are approximately collapsed together plotting $N_c$ as function of the 
rescaled quantity $(\sigma \alpha_c)^{-1/2}$. A continuous phase transition is identified, above which the system is shown to asymptotically 
fragment in several opinion clusters. The proposed scaling is sound in term of its psychological interpretation.  When $\alpha_c$ gets small 
the barrier in affinity fades off and the agents update their beliefs virtually at any encounter. The imposed selection rule drives a rapid 
evolution towards an asymptotic fragmented state, by favouring the interaction of candidates that share a similar view ($\Delta O_{ij}$ small). 
This tendency can be counter-balanced by adequately enhancing the social mixing, which in turn amounts to increase the value of 
$\sigma \propto \alpha_c^{-1}$. On the other hand, for large values of $\alpha_c$ the system is initially experiencing a  
lethargic regime, due to the hypothesized thresholding mechanism. Agents' opinions are therefore temporarily freezed to their initial values, 
while occasional encounters contribute to increase the degree of coehesion (synchronization) of the community. As the affinity grows, the social 
metric $D_{ij}$ becomes
less sensitive to $\Delta O_{ij}$ and the system naturally flows towards an ordered (single-clustered) configuration.
Notice that our system 
displays intriguing similarities with granular media, that have been shown to develop analogous self-organization features. This entails
the possibility of addressing the analysis of the observed structures within a purely statistical mechanics setting,  where the balance between    
competing effects is esplicitly modelled \cite{BenNaimTarzia}.

\begin{figure}[htbp]
\centering
\includegraphics[width=8cm]{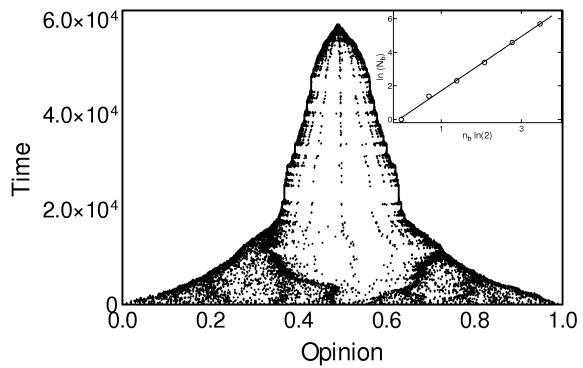}
\caption{Main plot: typical evolution in the mono-clustered phase. Inset: $N_b$ vs. $l=2^{-n_b}$ in log-log scale. For the choice of the parameters refer to the 
caption of Figure \ref{fig2}}
\label{fig3}
\end{figure}

\begin{figure}[htbp]
\centering
\includegraphics[width=7cm]{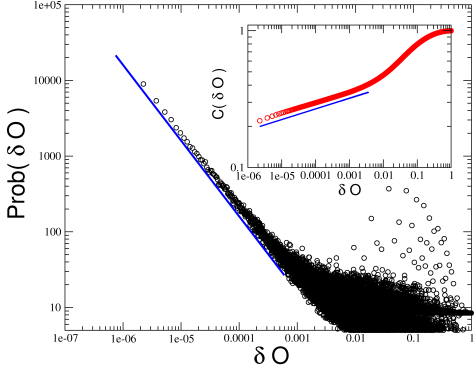}
\caption{Main frame: Histogram of $\delta O$, as follows from the numerics (N=100, averaged over 1000 independent realizations), 
plotted in log-log scale (symbols). The solid line is a guide for the eye. Inset: Cumulative distribution of the differences 
$\delta O$, in log-log scale.}
\label{fig4}
\end{figure}

Aiming at further characterising the process of convergence we have
also analysed  the following indicators: the fractal
  dimension of the orbits topology and the distribution of opinion differences. First, we focused on the single-clustered phase 
(main plot in Figure \ref{fig3}) and calculated the fractal dimension in the $(O,t)$ plane, a parameter that relates to the geometrical aspects of the 
dynamical evolution. A standard  box-counting algorithm is applied, which consists in partitioning the plan in small cells 
and identifying the boxes visited by the system trajectory.  In this specific case, the space $(O,t)$ is mapped into $[0,1]\times [0,1]$, and covered with
a uniform distribution of squares of linear size $l$. The number of filled box $N_b$ is registered and the measure repeated for different choices of 
$l$. In particular we set $l=2^{-n_b}$, where $n_b=1,2,..$.  For each
$n_b$, $N_b$ is plotted vs. $l$, in log-log scale (see inset of Figure
\ref{fig3}): A power-law  
decay is detected, whose exponent $\gamma \simeq 1.57$, quantifies
the fractal dimension. The orbits are also analyzed in the
multi-clustered regime and similar conclusion are drawn. In addition,   
every single cluster is isolated and studied according to the above
procedure, leading to an almost identical $\gamma$. 
In Figure \ref{fig4} we also report the probability distribution
function of $\delta O = |O_i^{t+1}-O_i^t|$. 
$\delta O$ measures the rate of change of individuals' opinion.  A power-law behaviour is found, an additional sign of system's criticality.  

\begin{figure}[htbp]
\vspace*{3.5em}
\centering
\includegraphics[width=7cm]{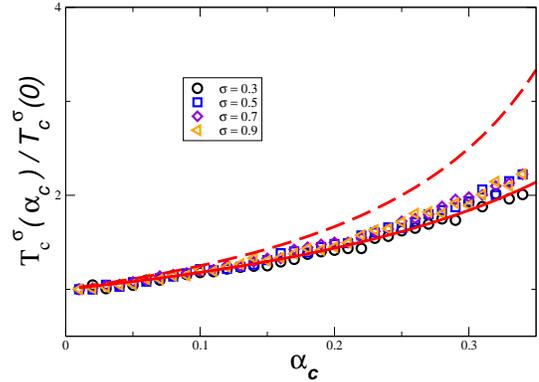}
\caption{Rescaled convergence time $T_c^\sigma(\alpha_c)/T_c^\sigma(0)$ is plotted as function of $\alpha_c$. Different symbols refer to
different values of the social temperature $\sigma$, see legend. The dashed line stands for the theoretical prediction (\ref{theoretical_estimate}).
The solid line is a numerical fit based on equation (\ref{theoretical_estimate}), where $\alpha_{max}$ is replaced by the effective value
$\alpha_{eff} = 0.66$ (see main text for further details).}
\label{fig5}
\end{figure}

Finally, working in the relevant mono-clustered regime, we also performed a dedicated campaign of simulations to estimate the convergence time, 
$T_c^\sigma(\alpha_c)$, i.e. the time needed to
  completely form the cluster under scrutiny. The experiments are conducted fixing the social temperature $\sigma$, and allowing $\alpha_c$ to span 
the interval $[0,\alpha_{max}]$, where $\alpha_{max} = \max_{i,j} \alpha_{ij}^0$  In Figure \ref{fig5} the rescaled
convergence time $T_c^\sigma(\alpha_c)/T_c^\sigma(0)$ is plotted as function of $\alpha_c$, for various choices of $\sigma$. All the different curves nicely 
collapse together, revealing an interesting positive correlation between the relative convergence time and the threshold $\alpha_c$. Again, this finding is 
certainly bound to reality: when $\alpha_c$ increases, individuals stick more rigidly to their opinion and changes happen only when encounters among neighbours 
occur. Instead, when reducing $\alpha_c$ large jumps in opinion are allowed which dynamically translate in a more effective mixing, hence faster 
convergence. To make this argument more rigorous, introduce $\mu' =
\mu [\tanh (\beta_1 (\alpha^t_{ij} -\alpha_c))+1]/2$. A reduced
dynamical formulation can obtained  
by averaging out the dependence on $\alpha_{i,j}$ in (\ref{opinion}),
thus formally  decoupling it from eq. (\ref{alpha}). This is
accomplished, at fixed $i$,  
as follows:
\begin{eqnarray}
<\mu '> &=& \mu \int \Gamma_1 (\alpha_{ij}^{t}) f_t(\alpha_{ij}^{t}) d \alpha_{ij}^{t} \\ \nonumber
&\simeq& \mu \int_0^{\alpha_{max}} \Gamma_1 (\alpha_{ij}^{0}) f_0(\alpha_{ij}^{0}) d \alpha_{ij}^{0} \\ \nonumber
&\simeq&  \mu \frac{\alpha_{max} - \alpha_c}{\alpha_{max}} \nonumber
\label{decoupl}
\end{eqnarray}
where in the last passage we made use of the fact that $\beta_1 \rightarrow \infty$ and $f_0(\alpha_{ij}^{0})=1/\alpha_{max}$ as it follows from the normalisation condition. 
The function $f_t(\cdot)$ (resp. $f_0(\cdot)$ ) represents the affinity distribution of agents $j$ versus $i$, at time $t$ (resp. at time zero). Within this simplified scenario, the time of convergence 
scales as $1/<\mu '>$ \cite{propaganda} and therefore expression (\ref{decoupl}) immediately yields to:
\begin{equation}
\frac{T_c^\sigma(\alpha_c)}{T_c^\sigma(0)} = \frac{\alpha_{max}}{\alpha_{max} - \alpha_c}
\label{theoretical_estimate}
\end{equation}
Relation (\ref{theoretical_estimate}) is reported  in Figure \ref{fig5} (dashed line) and shown to approximately reproduce the 
observed functional dependence. A good agreement with direct simulations is found for small $\alpha_c$. It however progressively 
deteriorates for larger $\alpha_c$, due to non-linear contributions. The latter can be incorporated into our scheme by replacing $\alpha_{max}$ in eq. 
(\ref{theoretical_estimate}) with an effective  value $\alpha_{eff}$, to be determined via numerical fit (solid line in Figure \ref{fig5}).  
Such a value accounts for the system tendency to populate the complementary  domain $1-\alpha_{max}$ and results in an excellent agreement with the simulated data.

In this Letter we introduced a new model for studying the process of opinion formation. This novel interpretative framework allows us to
account for the affinity, an effect of paramount importance in real social
systems. The model here proposed captures the essence of the 
cognitive dissonance theory, a psychological construction elaborated by L. Festinger in the late 50s. Numerical investigations are carried on 
and reveal the presence of a phase transition between an ordered (single clustered) and a disordered (multi-clustered) phase. Evidence of 
critical behaviours is provided, and the role of different parameters elucidated. We firmly believe that our formulation represents a leap forward 
in social system modelling, thus opening up new perspectives to reinforce the ideal bridge with the scattered psychology community.


\begin{thebibliography}{99}
\bibitem{opinion} D. Stauffer and M. Sashimi, Physics A {\bf 364}, 537, (2006); A. Pluchino et al. Eur. Phys.J. B {\bf 50}, 169 (2006); 
A. Baronchelli et. al. preprint cond-mat/0611717.
\bibitem{Sznajd} K. Sznajd-Weron, J. Sznajd, Int. J. Mod. Phys. C {\bf 11}, 1157 (2000).
\bibitem{Deffuant} G. Deffuant et al. Adv. Compl. Syst. {\bf 3}, 87 (2000).
\bibitem{Nowak} Nowak et al., Developmental Review, {\bf 25}, Issues 3-4, (2005), 351--385.
\bibitem{festinger} L. Festinger and J.M. Carlsmith, Journal of Abnormal and Social Psychology, {\bf 58}, 203-210 (1959).
\bibitem{guarino} S. Ciliberto, A. Guarino, R. Scorretti, Physica D {\bf 158}, 83-104 (2001).
\bibitem{BenNaimTarzia} E. Ben-Naim, P.L. Krapivsky, Journ. Phys. A: Math Gen. {\bf 38}, 417-423 (2005); M. Tarzia et al., Phys. Rev. Lett., {\bf 95}  078001 (2005).
\bibitem{propaganda} T. Carletti et al., Europhys. Lett.  {\bf 74}, 222 (2006). 
\end{thebibliography}
\end{document}